\documentclass[12pt]{article}
\usepackage{graphicx}
\usepackage{lineno}

\newcommand\pubnumber{CIPANP2018-Masubuchi}
\newcommand\pubdate{\today}

\def\icepp{International Center for Elementary Particle Physics\\
The University of Tokyo, Japan}
\def\support{ on behalf of the ATLAS collaboration}

\newcommand\pt{$p_{\mathrm{T}}$}

\newcommand\bbbar{$b\bar{b}$}
\newcommand\ifb{fb$^{-1}$}
\newcommand\MET{$E_{\mathrm{T}}^{\mathrm{miss}}$}
\newcommand\ptv{$p_{\mathrm{T}}^{V}$}
\newcommand\mbb{$m_{bb}$}
\newcommand\mcc{$m_{cc}$}
\def\Title#1{\begin{center} {\Large #1 } \end{center}}
\def\Author#1{\begin{center}{ \sc #1} \end{center}}
\def\Address#1{\begin{center}{ \it #1} \end{center}}

\newcommand\pubblock{\rightline{\begin{tabular}{l} \pubnumber\\
         \pubdate  \end{tabular}}}
\newenvironment{Abstract}{\begin{quotation}  }{\end{quotation}}
\newenvironment{Presented}{\begin{quotation} \begin{center} 
             PRESENTED AT\end{center}\bigskip 
      \begin{center}\begin{large}}{\end{large}\end{center} \end{quotation}}

\def\beq{\begin{equation}}
\def\eeq#1{\label{#1}\end{equation}}
\def\eeqn{\end{equation}}

\def\beqa{\begin{eqnarray}}
\def\eeqa#1{\label{#1}\end{eqnarray}}
\def\eeqan{\end{eqnarray}}

\let\bar=\overbar

\def\Dslash{\not{\hbox{\kern-4pt $D$}}}
\def\dslash{\not{\hbox{\kern-2pt $\del$}}}

\def\msb{{\bar{\ssstyle M \kern -1pt S}}}

\begin{document}
\begin{titlepage}
\pubblock

\vfill
\Title{Measurements and Searches of Higgs boson \\decays to two fermions}
\vfill
\Author{ Tatsuya Masubuchi \\\support \footnote{Copyright 2018 CERN for the benefit of the ATLAS Collaboration. CC-BY-4.0 license.}}
\Address{\icepp}
\vfill
\begin{Abstract}
The interaction of the Higgs boson to fermion is most fragile part in the Standard model. The Higgs decay to ditau ($H\rightarrow \tau\tau$) was observed and established in the LHC Run 1 data. This report presents recent measurements and searches of Higgs boson decay to two fermions. In particular this report focuses on results of the Higgs boson decays to third-generation ($H\rightarrow b\bar{b}$) and second-generation ($H\rightarrow c\bar{c}$, $H\rightarrow \mu\mu$) fermions using the LHC Run 2 data collected by the ATLAS experiment at the LHC. 
\end{Abstract}
\vfill
\begin{Presented}
Thirteenth Conference on the Intersections of Particle and Nuclear Physics (CIPANP2018)\\
Palm Springs, CA, USA,  May 29 - June 3, 2018
\end{Presented}
\vfill
\end{titlepage}
\def\thefootnote{\fnsymbol{footnote}}
\setcounter{footnote}{0}

\section{Introduction}
A Higgs boson was discovered in 2012 by the ATLAS and CMS collaborations~\cite{ATLASdiscovery, CMSdiscovery}. The properties of the discovered Higgs boson has been predominantly measured using bosonic decay mode, $H\rightarrow \gamma\gamma$, $H\rightarrow ZZ$ and $H\rightarrow WW$ in LHC Run 1 and Run 2~\cite{ATLASComb, ATLASSpinComb, ATLASMass}. The measured properties are compatible with the predictions of the Standard Model (SM) within uncertainties. However, a great mystery on the interaction (Yukawa coupling) between Higgs and fermions still remains in the Higgs sector of the SM. It is important to experimentally measure Yukawa coupling for each fermion and to test if the coupling is proportional to fermion mass to solve the origin of the fermion mass and generation.
Direct evidences for the coupling of the Higgs boson to fermions were established via the decay to $\tau$-leptons through the combination of ATLAS and CMS Run 1~\cite{ATLASCMSComb} and recently the Higgs production associated with top quarks~\cite{CMSttH}. 
Table~\ref{tbl:BR} shows the branching ratio of Higgs decay to fermion~\cite{YR4}. The coupling of Higgs boson to bottom quark is not observed yet although the decay branching ratio is dominant at $m_{H}=125$~GeV. In LHC Run 2 the observation of $H\rightarrow b\bar{b}$ mode is one of big milestones in Higgs physics of the ATLAS experiment~\cite{ATLAS}. 

\begin{table}[h]
  \begin{center}
    \begin{tabular}{ccccc}  
\hline
$m_{H}=125$ GeV       & $b\bar{b}$ &  $\tau\tau$ & $c\bar{c}$ & $\mu\mu$ \\ \hline 
Branching Ratio & 58.2\% & 6.27\% & 2.89\% & 0.023\% \\ \hline
    \end{tabular}
    \caption{The branching ratio of major Higgs decays to fermions at $m_{H}$=125 GeV. }
    \label{tbl:BR}
  \end{center}
\end{table}

Various theories beyond the Standard model (BSM) predict the deviation of Yukawa couplings. The precise measurement of the coupling between the Higgs and fermions might unveil the BSM and determine the physics model since advocated BSM models predict different deviation patterns on the Yukawa coupling. 

\section{Higgs decay to bottom quark pair }

The decay of the SM Higgs boson to pairs of $b$-quarks is expected to have a branching ratio of 58\% at $m_{H} = 125$\ GeV, the largest among all decay modes. At the LHC, the enormous backgrounds arising from multi-jet production make an inclusive search extremely challenging. The most sensitive production modes for $H\rightarrow b\bar{b}$ search are those where the Higgs boson is produced in association with a $W$ or $Z$ boson. Their leptonic decay leads to clean signatures that can be triggered by the high-\pt\ electron/muon or missing transverse energy (\MET), and can reject multi-jet background. 

Another new approach using vector boson fusion (VBF) production has been attempted. In order to enhance signal-to-background ratio in the VBF production mode, a photon radiated from an internal $W^{\pm}$ boson or from an incoming or outgoing quark is required. 

\subsection{Evidence for the $H\rightarrow b\bar{b}$ decay using Higgs boson production in association with a $W$ or $Z$ boson}
\label{sec:VHbb}

A search for the decay of Higgs boson to a pair of $b$-quarks using the Higgs boson production in association with a $W$ or $Z$ boson is reported~\cite{VHbb}. The data used in this analysis were collected at a center-of-mass energy of 13 TeV during 2015 and 2016 running period, and correspond to an integral luminosity of 36.1~\ifb. Three main signatures are explored, $ZH\rightarrow \nu\nu b\bar{b}$, $WH\rightarrow \ell\nu b\bar{b}$ and $ZH\rightarrow \ell\ell b\bar{b}$. These channels are referred to as 0-,1- and 2-lepton channels, based on the number of selected charged leptons. All events are required to have at least two jets, and exactly two jets must pass the $b$-tagging requirement. The event selection has been optimized for each lepton channel. Anti-QCD selections using the angular correlations between \MET\ and (di)jet-systems are applied to reduce further multi-jet background to negligible level in the 0 lepton channel. Dedicated identification and isolation requirements are required in the charged lepton to reduce multi-jet background in the 1 lepton channel. The reconstructed dilepton mass should be consistent with the $Z$ boson mass in the 2 lepton channel. 
Selected events are categorized according to the reconstructed vector boson's transverse momentum \ptv\ to enhance signal-to-background ratio. In the 0- and 1-lepton channels a signal region is defined, with \ptv $> 150$\ GeV. In the 2-lepton channel two regions are considered, 75\ GeV $<$ \ptv $<$ 150 GeV and \ptv $>$ 150 GeV as shown in Figure~\ref{fig:pTV_Mbb}. Events are further split into two categories according to the jet multiplicity. In 0- and 1-lepton channels, events are considered with exactly two or three jets. In the 2-lepton channel, events are considered with exactly two or three or more jets. 

The reconstructed dijet mass resolution is one of key ingredients to improve search sensitivity in this analysis. Customized $b$-jet energy corrections are applied in addition to general jet energy scale correction. Figure~\ref{fig:pTV_Mbb} shows \mbb\ distribution with various $b$-jet energy corrections. In particular, the $ZH\rightarrow \ell\ell b\bar{b}$ event kinematics can be used to constrain $b$-jet energy, with a kinematic likelihood due to no intrinsic missing energy in the 2 lepton channel. This provides about 40\% improvement of the dijet mass resolution. 

\begin{figure}[htb]
\centering
\includegraphics[width=6.7cm]{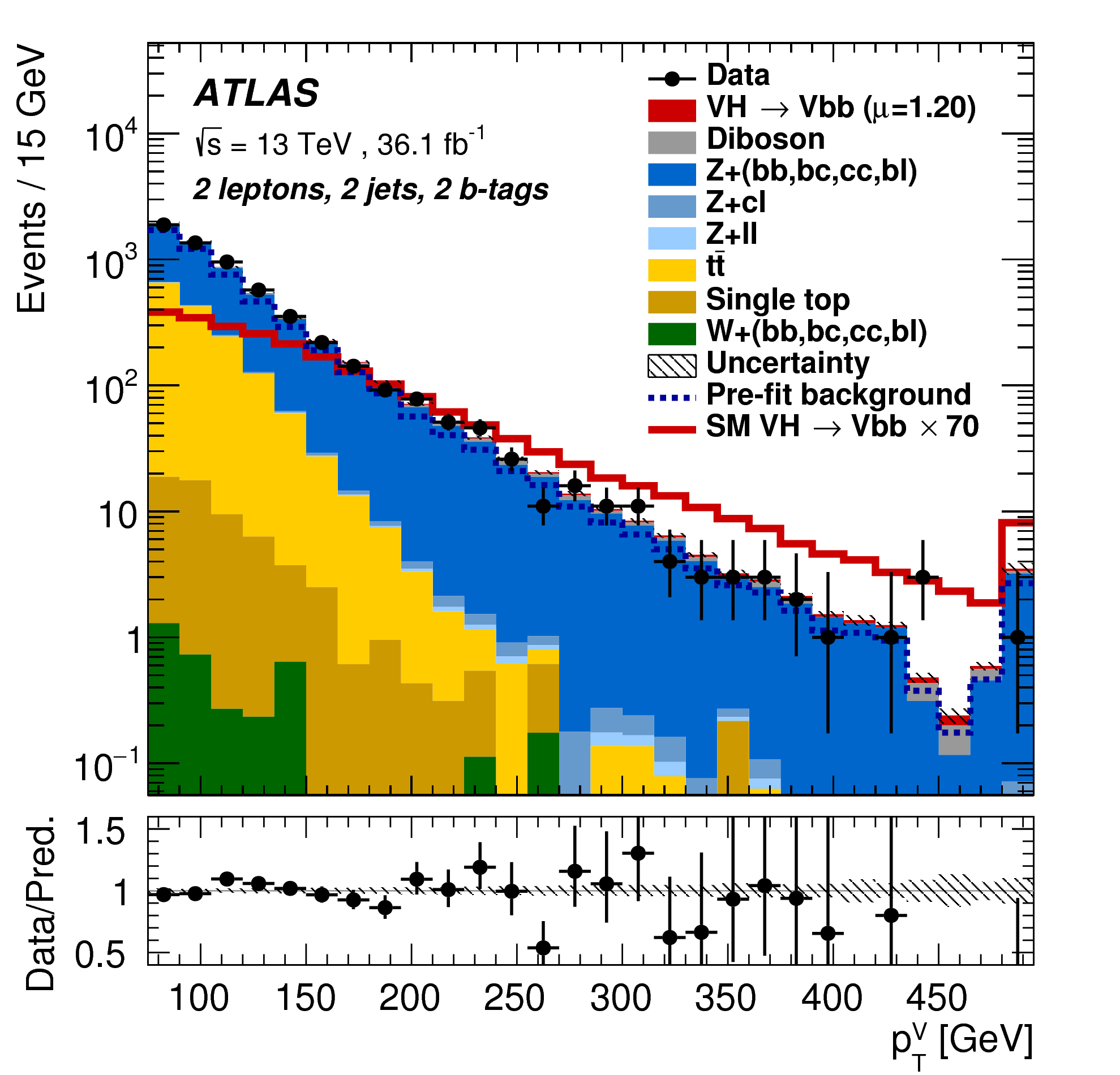}
\includegraphics[width=8.3cm]{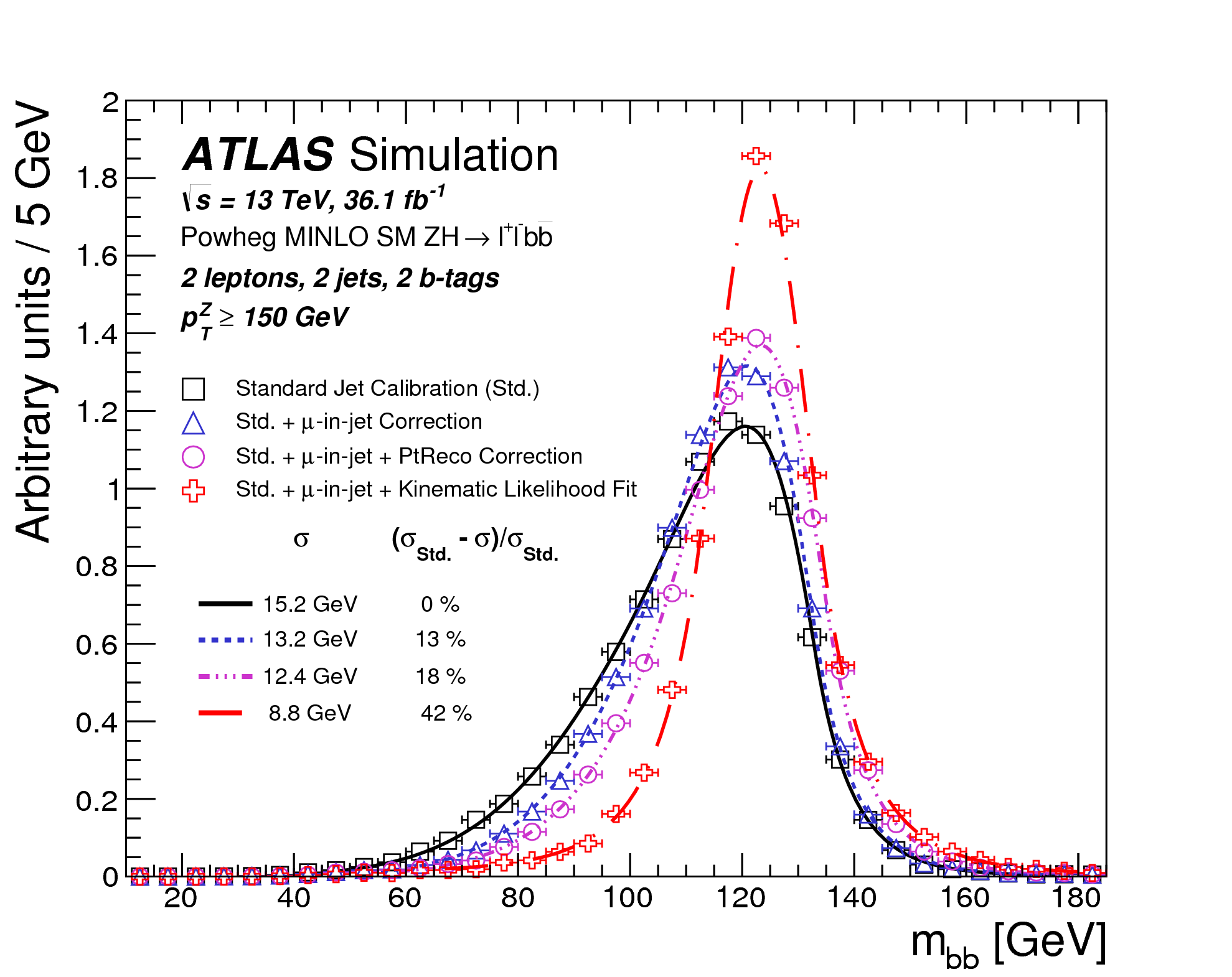}
\caption{\ptv\ distribution in 2 lepton, 2-jet signal region (left). Comparison of the \mbb\ distributions as additional correction are applied to the jet energy scale for the 2-lepton channel in the 2-jet and \ptv $>$ 150 GeV region (right)~\cite{VHbb}.}
\label{fig:pTV_Mbb}
\end{figure}

Another important ingredient is multivariate analysis. Multivariate discriminants making use of boosted decision trees (BDTs) are constructed, trained and evaluated in each lepton channel and analysis region separately.  In total eight signal regions and two $W$+jets control regions(CRs), which enhances $W$+jets purity in the selection of low \mbb\ and high $m_{\mathrm{top}}$\ in the 1 lepton channel and four top control regions requiring different lepton flavor ($e\mu$) in the 2 lepton channel are fit simultaneously. The BDT output distributions are used in the signal region, while one bin distributions in $W$+jets CRs to determine background normalization and \mbb\ shape in top control regions are used. Experimental and background modeling uncertainties are assigned in each signal region and control region accordingly. Figure~\ref{fig:BDToutput} shows fitted BDT output distributions.

\begin{figure}[htb]
\centering
\includegraphics[width=4.5cm]{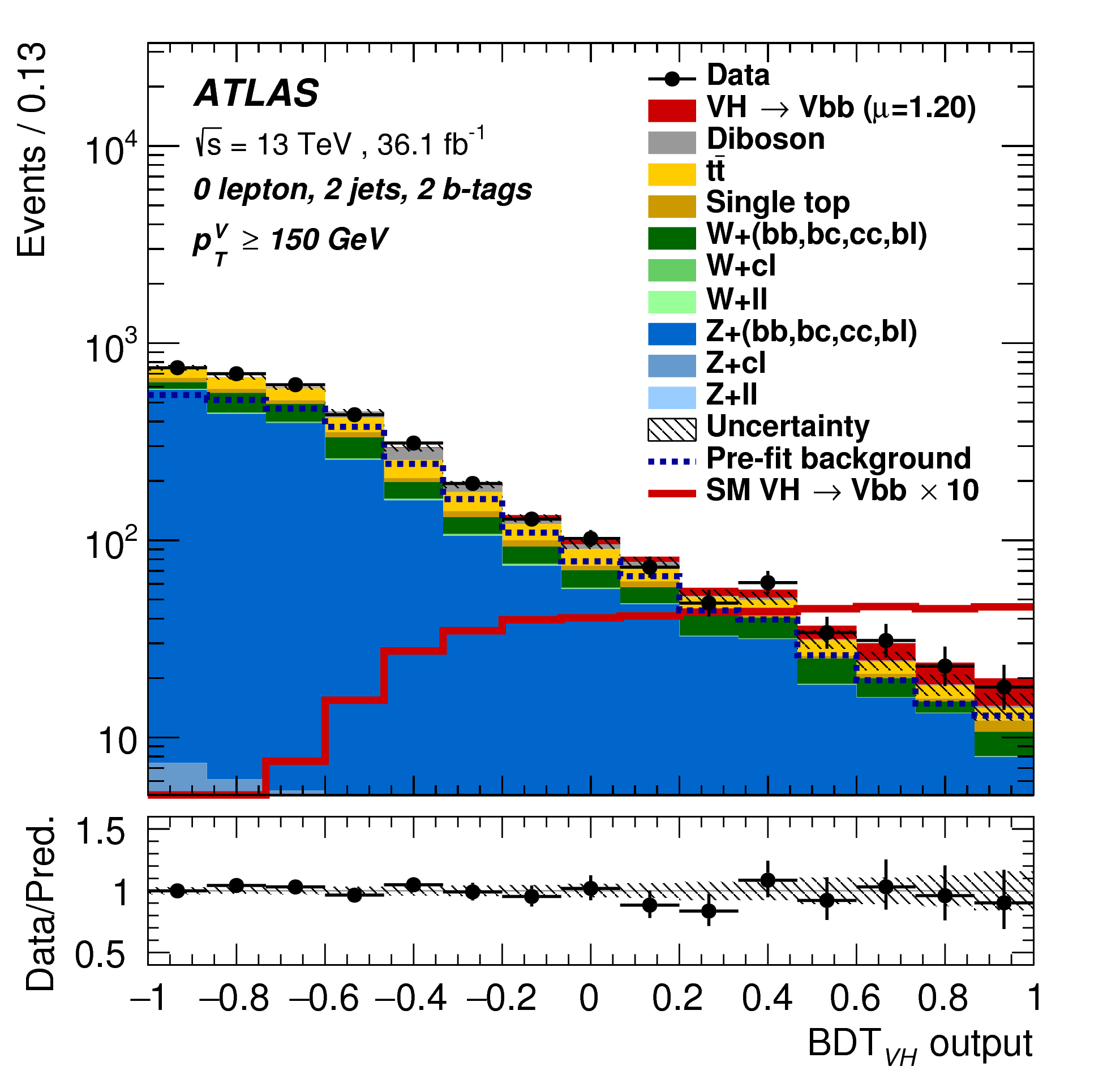}
\includegraphics[width=4.5cm]{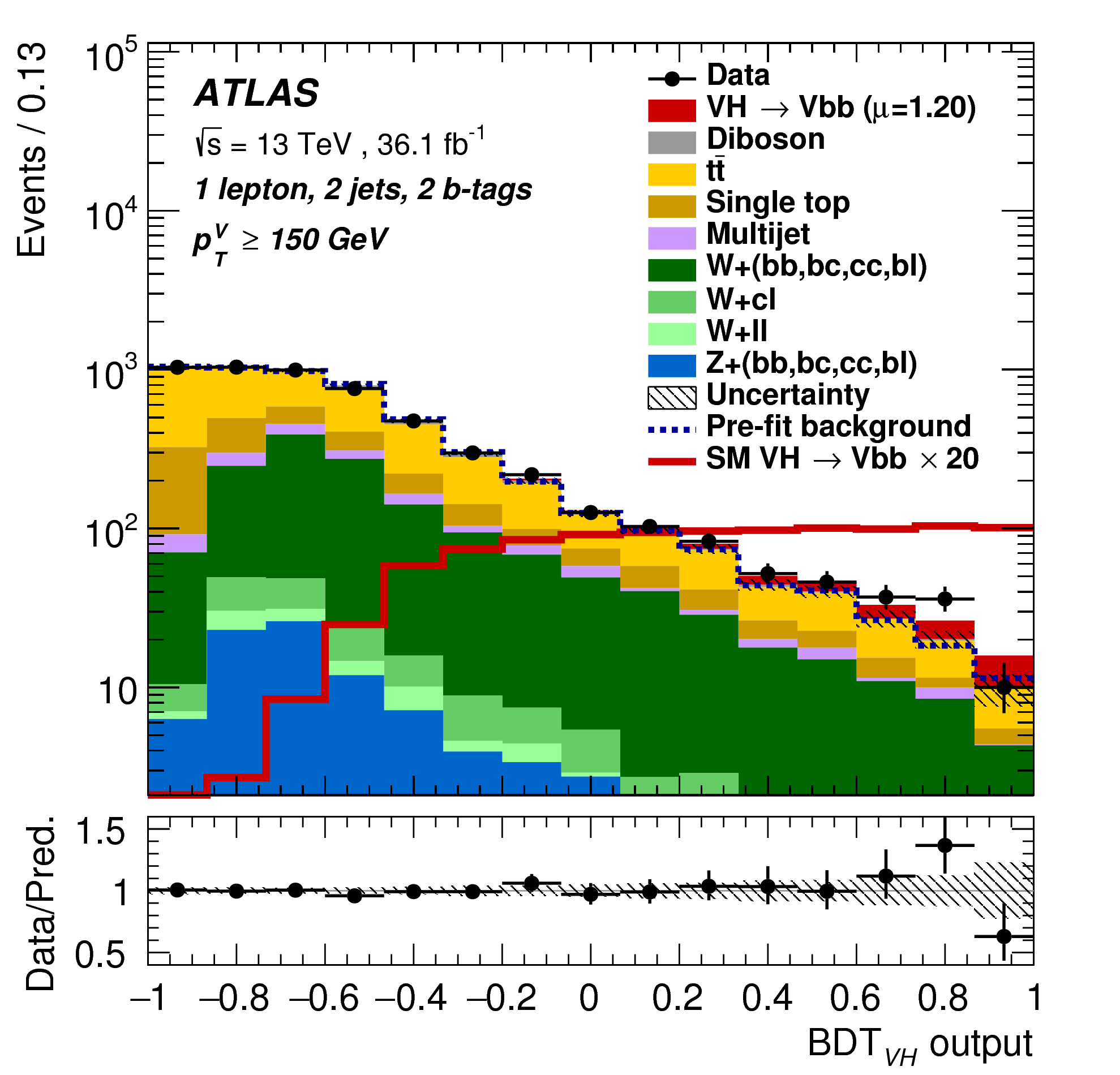}
\includegraphics[width=4.5cm]{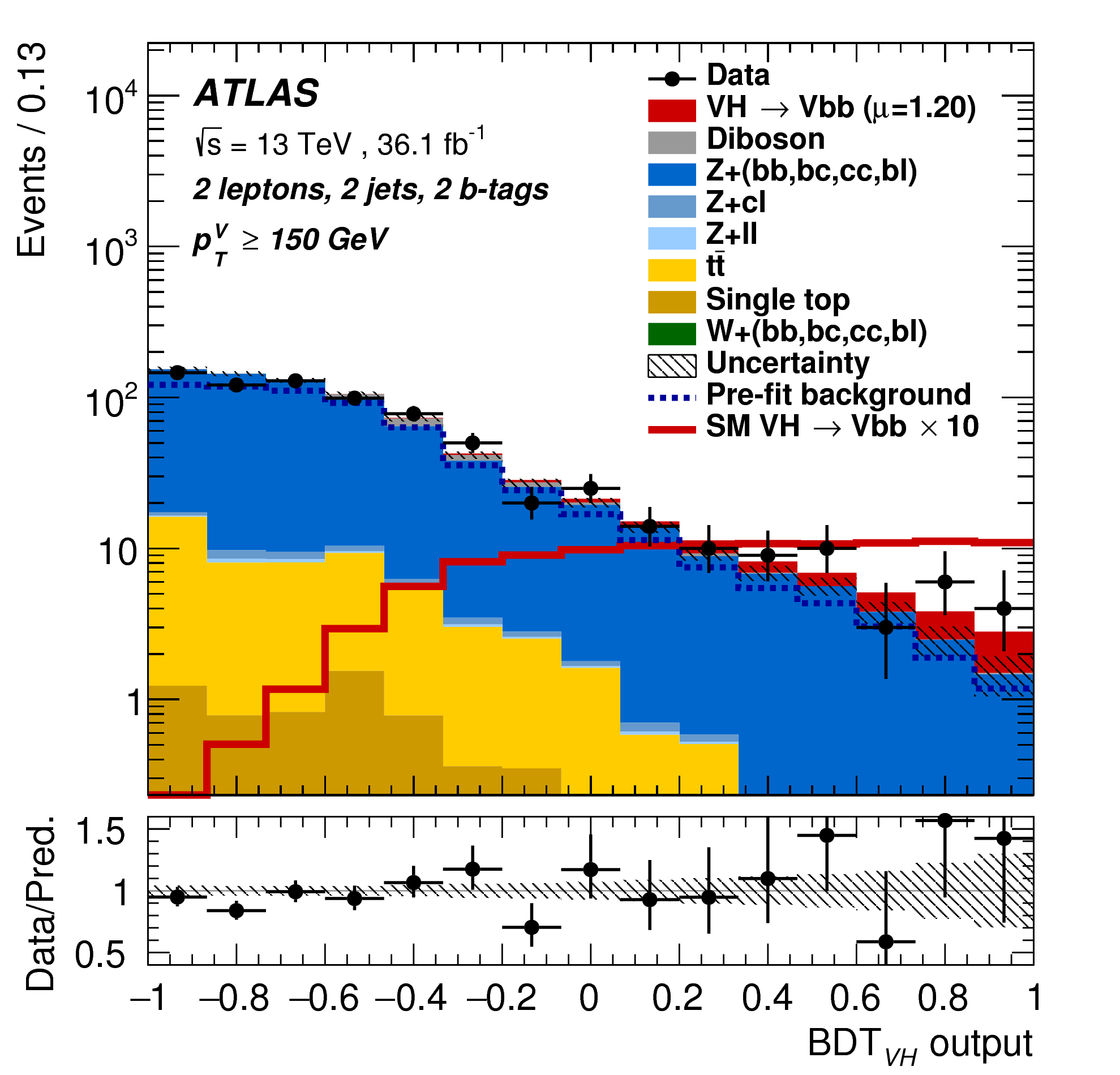}
\caption{The BDT$_{VH}$ output distribution in the 0-lepton (left), 1-lepton (middle) and 2-lepton (right) for the 2-jet and \ptv $>$ 150 GeV region~\cite{VHbb}. }
\label{fig:BDToutput}
\end{figure}

This complicated background modeling scheme has been validated with diboson ($VZ\rightarrow b\bar{b}$) background. The BDTs are re-trained with the diboson process as signal. The observed signal strength is $\mu_{VZ}^{b\bar{b}} = 1.11^{+0.25}_{-0.22}$, which is consistent with the SM. The observed significance is 5.8 standard deviations.
 
In the $VH\rightarrow b\bar{b}$ signal extraction a combined fit with Run 1 data~\cite{VHbbRun1} has been performed. The observed signal strength is $\mu_{VH}^{b\bar{b}} = 0.90^{+0.28}_{-0.26}$ as shown in Figure~\ref{fig:VHbbresult}. The observed significance is 3.6 standard deviations, to be compared to an expectation of 4.0 standard deviations. 

\begin{figure}[htb]
\centering
\includegraphics[width=8.3cm]{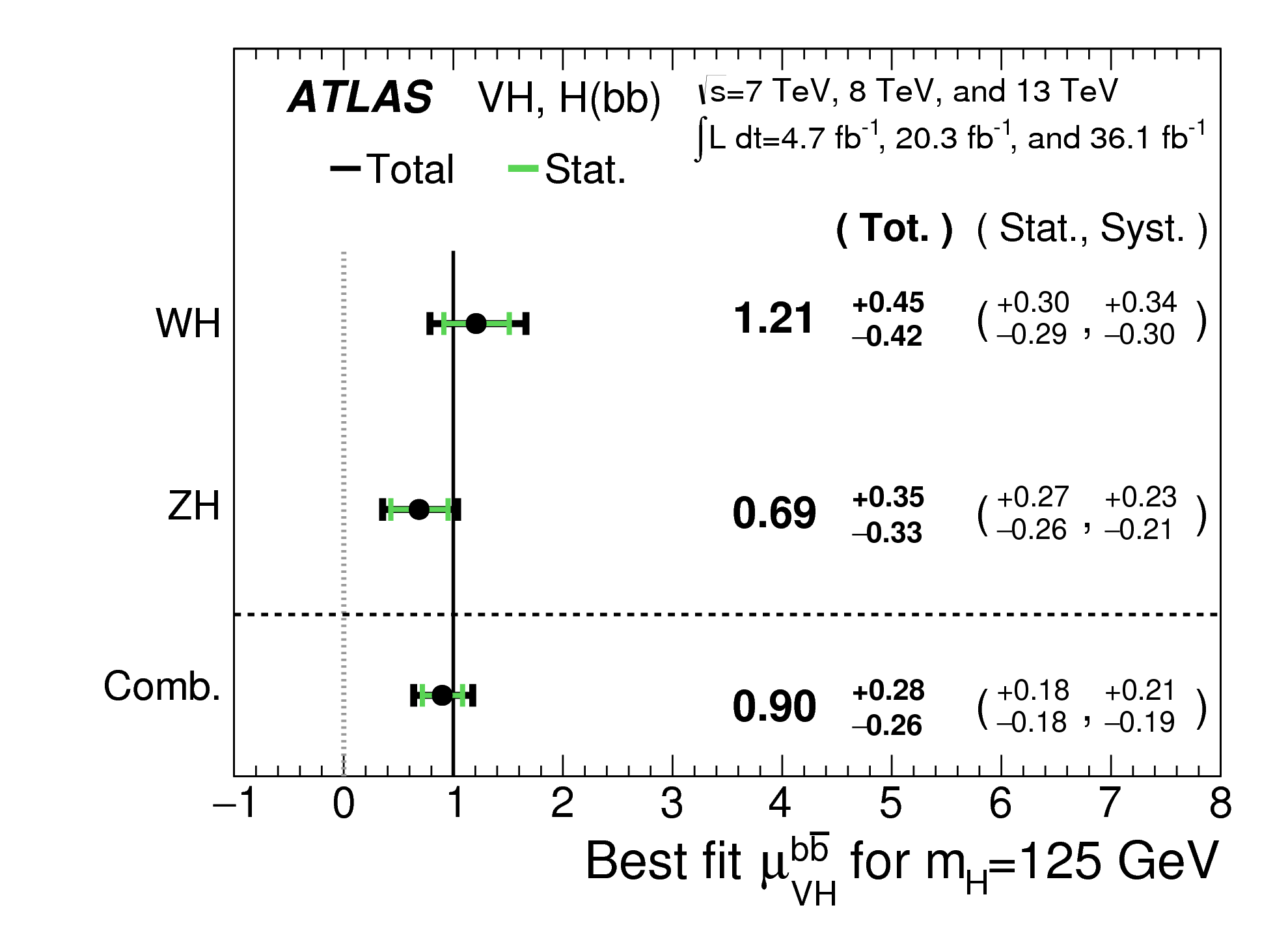}
\includegraphics[width=6.7cm]{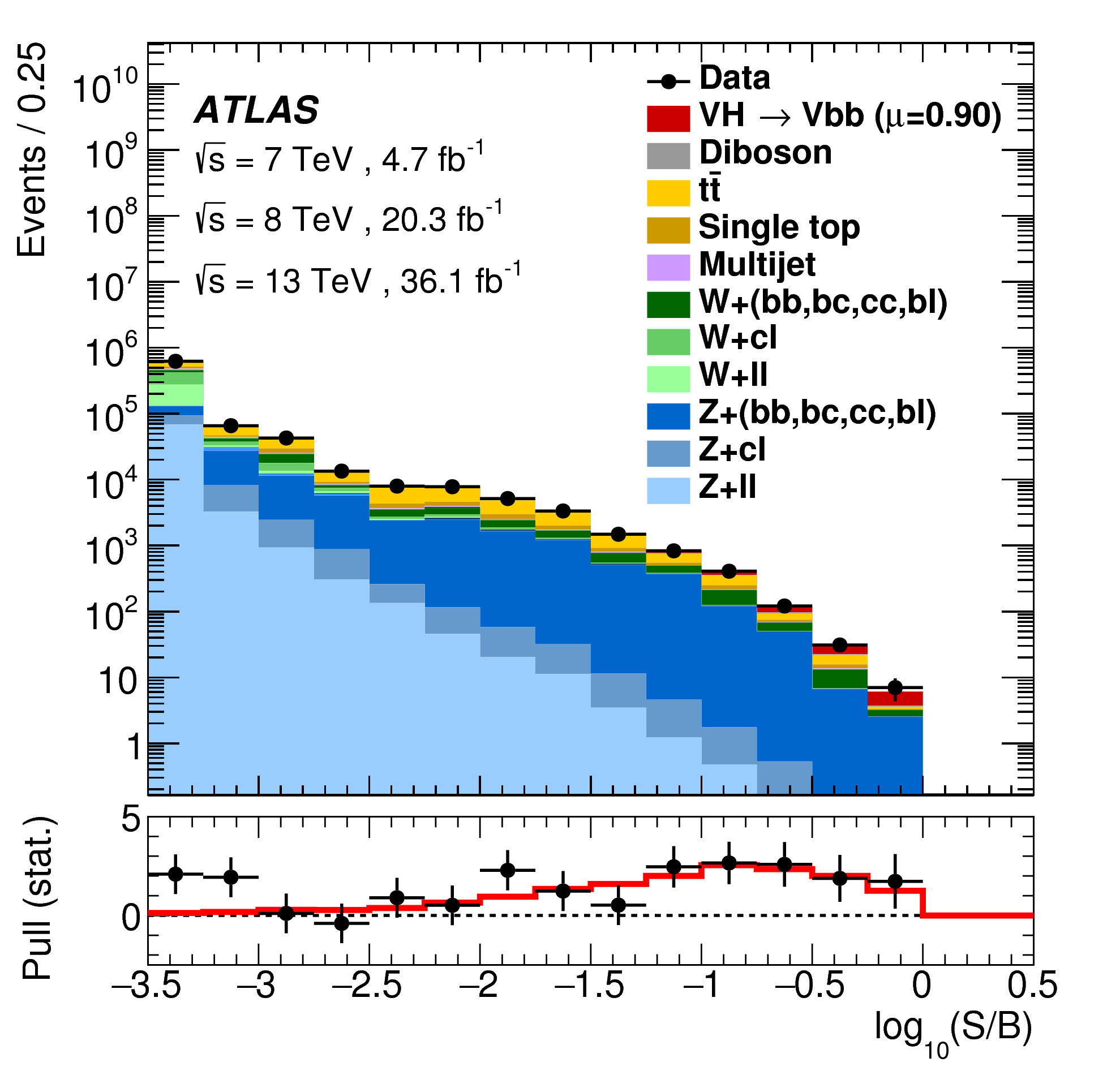}
\caption{The fitted signal strength parameter for the $WH$ and $ZH$ processes (left). Events yields as a function of $\log(S/B)$ for data, background and the Higgs boson signal (right). The Higgs boson signal contribution is scaled to $\mu=0.90$~\cite{VHbb}.}
\label{fig:VHbbresult}
\end{figure}

\subsection{Search for Higgs boson production via weak boson fusion and decaying $b\bar{b}$ with a high-energy photon}
\label{sec:VBFbb}

The VBF production mode for the Higgs decay to pair of $b$-quarks is challenging due to larger contributions from non-resonant $bbjj$ production and difficulties in triggering on low-\pt\ $b$-jets. A novel study that is requiring high-\pt\ photon~\cite{VBFbb} has been performed. The presence of high-\pt\ photon provides a clean signature and efficient triggering in the hadron collider. Furthermore, the background diagrams suppress photon emission from gluon line and destructive interference between diagrams with photon emission from the initial-state quark and the final-state quark further reduces the background cross section as shown in figure~\ref{fig:VBFgdiagram}. 

A dedicated trigger has been developed for VBF+photon analysis. The offline event selections require one photon with \pt\ $>$ 30 GeV and at least four jets with \pt\ $>$ 40 GeV and $|\eta|$ $<$ 4.5, and at least two of them must be $b$-tagged in the central detector region ($|\eta|<2.5$). If more than two $b$-tagged jets were found, jets with highest and second highest $b$-tagging discriminant output would be taken as a Higgs($b\bar{b}$) system in the candidate events. The \bbbar\ system must have \pt\ $>$ 80 GeV to reduce distortion of invariant mass distribution and the dijet mass reconstructed from non-Higgs jets should satisfy $m_{jj} > 800$ GeV. 

After offline event selection, a BDT discriminant is constructed from a set of variables based on VBF topology ($m_{jj}$ , $\Delta\eta_{jj}$, jet width to separate gluon jet from quark and so on). Figure~\ref{fig:VBFresult} shows a BDT output distribution. The BDT output is further split into three categories depending on the BDT score. The signal extraction is performed by fitting the \bbbar\ invariant mass distributions of three BDT categories. Resonant signals are modeled with Crystal ball functions and non-resonant backgrounds are modeled with a second-order polynomial function. 
Figure~\ref{fig:VBFresult} shows \mbb\ distribution in the highest purity signal region. No significant excess over background expectation is observed. An observed limit on signal strength is 4.0, to be compared with an expectation of $6.0^{+2.3}_{-1.7}$. 

\begin{figure}[htb]
\centering
\includegraphics[width=4.5cm]{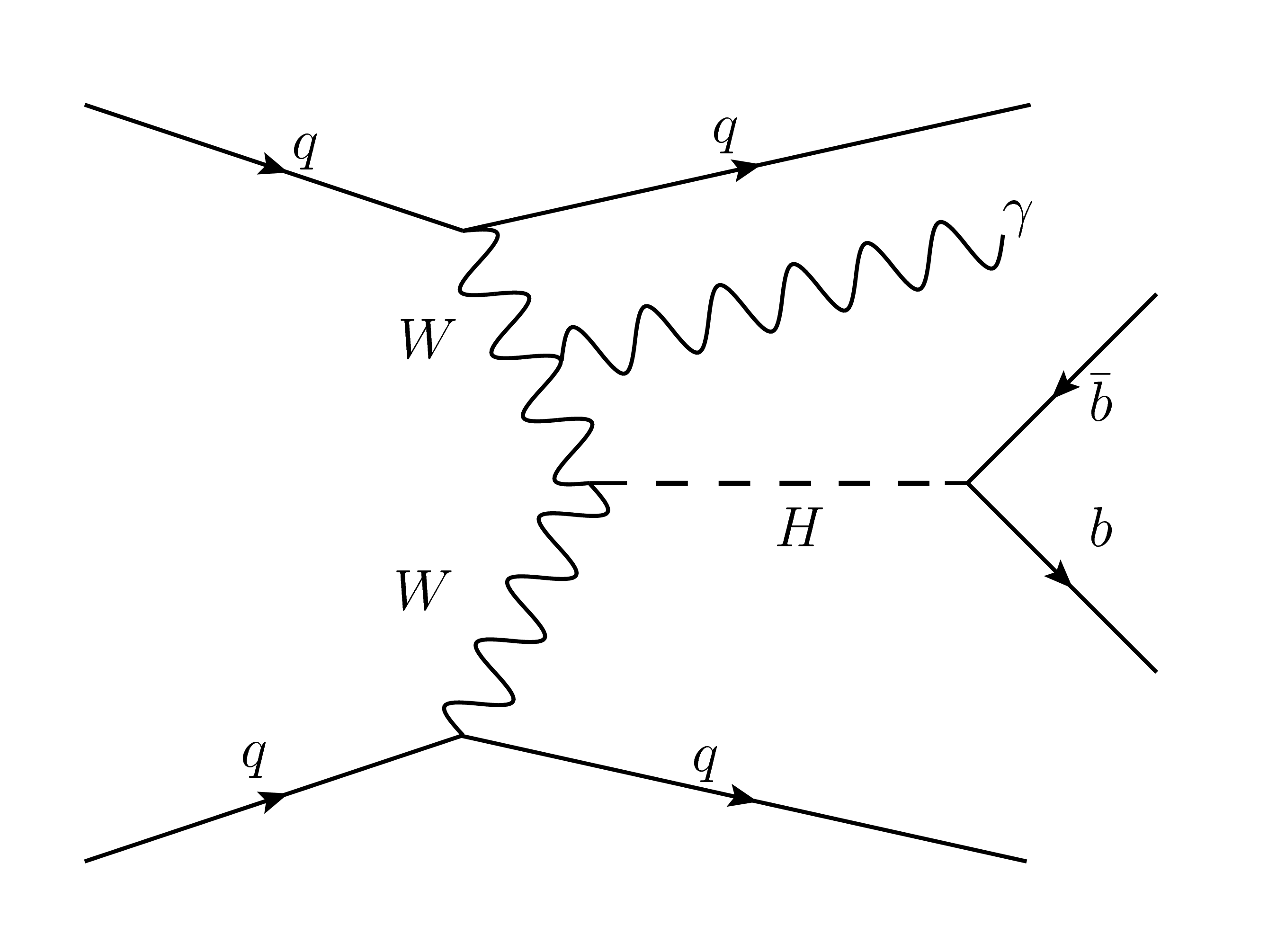}
\includegraphics[width=3.5cm]{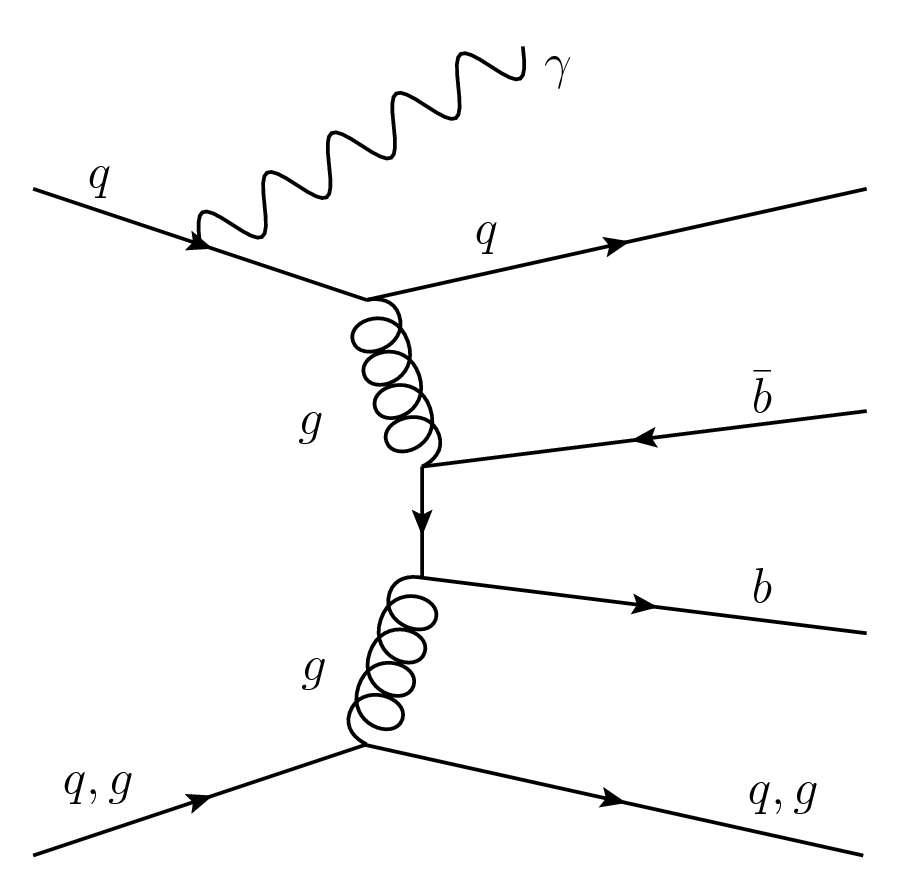}
\includegraphics[width=3.5cm]{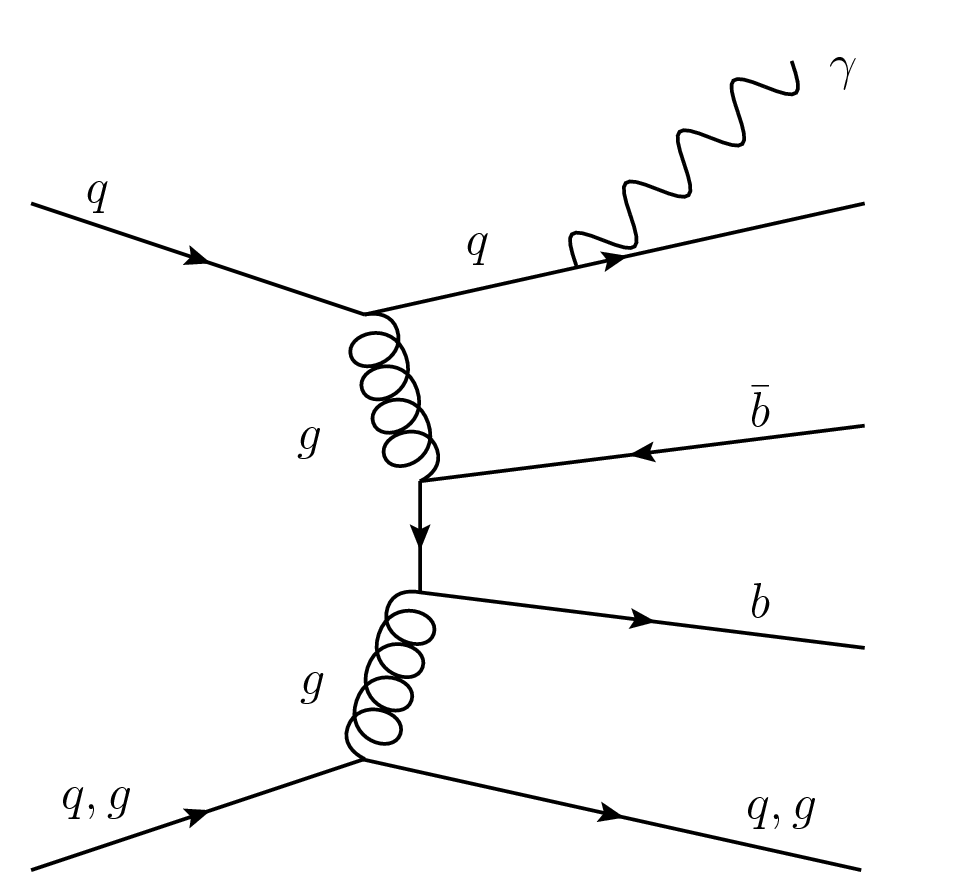}
\caption{Representative leading-order Feynman diagrams for Higgs boson production via vector boson fusion in association with a photon (left) and non-resonant b anti-b production in association with a photon and jets (middle, right)~\cite{VBFbb}.}
\label{fig:VBFgdiagram}
\end{figure}

\begin{figure}[htb]
\centering
\includegraphics[width=8cm]{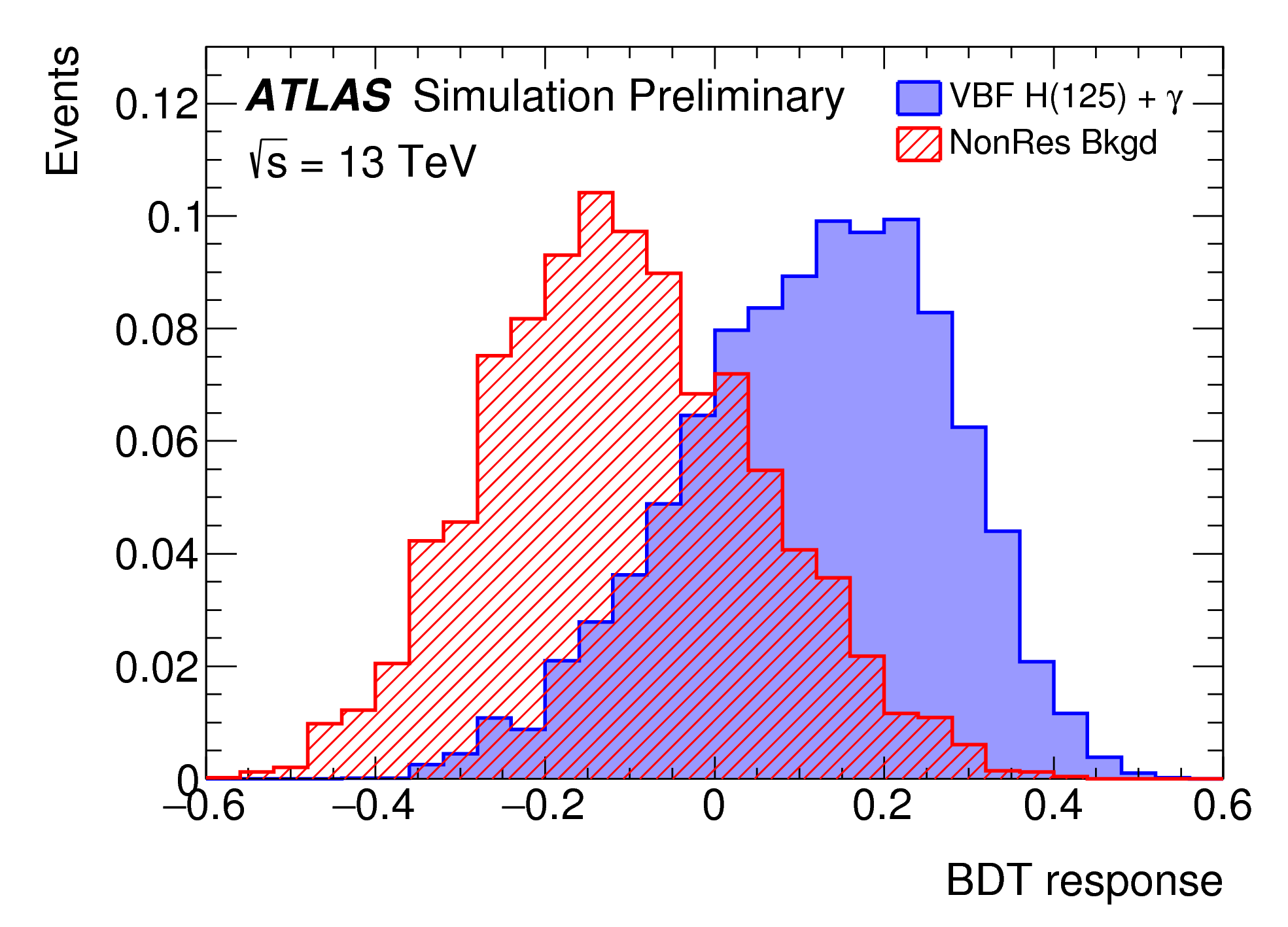}
\includegraphics[width=7cm]{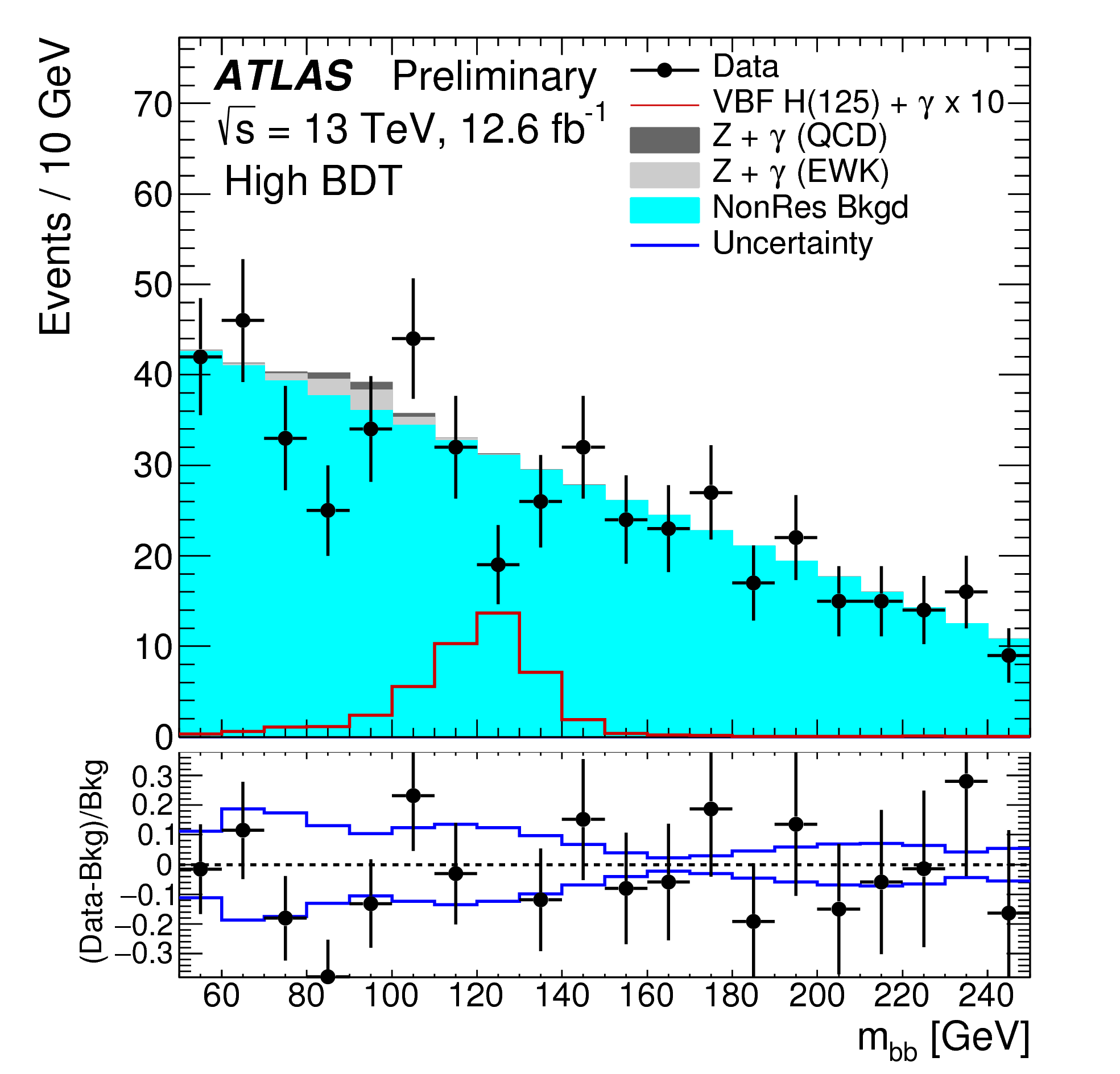}
\caption{BDT output distribution (left). \mbb\ distribution in the highest BDT region~\cite{VBFbb}. }
\label{fig:VBFresult}
\end{figure}

\section{Search for the Higgs boson decay to charm quark}
\label{sec:Hcc}
A direct search for the Higgs decaying to a pair of charm quarks is reported using the Higgs boson production in associated with a $Z$ boson, in the 2 lepton decay mode ($ZH\rightarrow \ell\ell c\bar{c}$)~\cite{VHcc}. The analysis strategy is similar to $ZH\rightarrow \ell\ell b\bar{b}$ analysis discussed in Section~\ref{sec:VHbb}. Direct $H\rightarrow c\bar{c}$ search is quite challenging because it is difficult to separate $c$-jet from $b$-jet and light-jet. However, dedicated flavor-tagging scheme is developed for $H\rightarrow c\bar{c}$ search. Figure~\ref{fig:VHccresult} shows an optimized $c$-tagging working point, which has 41\%, 25\%, 5\% $c$-tagging, $b$-tagging and light-tagging efficiency, respectively. Analysis requires one or two $c$-tagged jet. The final discriminant is fitted $m_{cc}$ distribution. Therefore, additional $\Delta R_{cc}$ selection is applied to reduce background. 
Figure~\ref{fig:VHccresult} shows $m_{cc}$ distribution for the signal region with \ptv $>$ 150 GeV. An observed upper limit on $\sigma(pp\rightarrow ZH)\times Br(H\rightarrow c\bar{c}$) is 2.7~pb, to be compared with an expectation of $3.9^{+2.1}_{-1.1}$~pb at the 95\% confidence level. The upper limit is approximately 100 times higher than the SM expectation in the current data. 

\begin{figure}[htb]
\centering
\includegraphics[width=7.5cm]{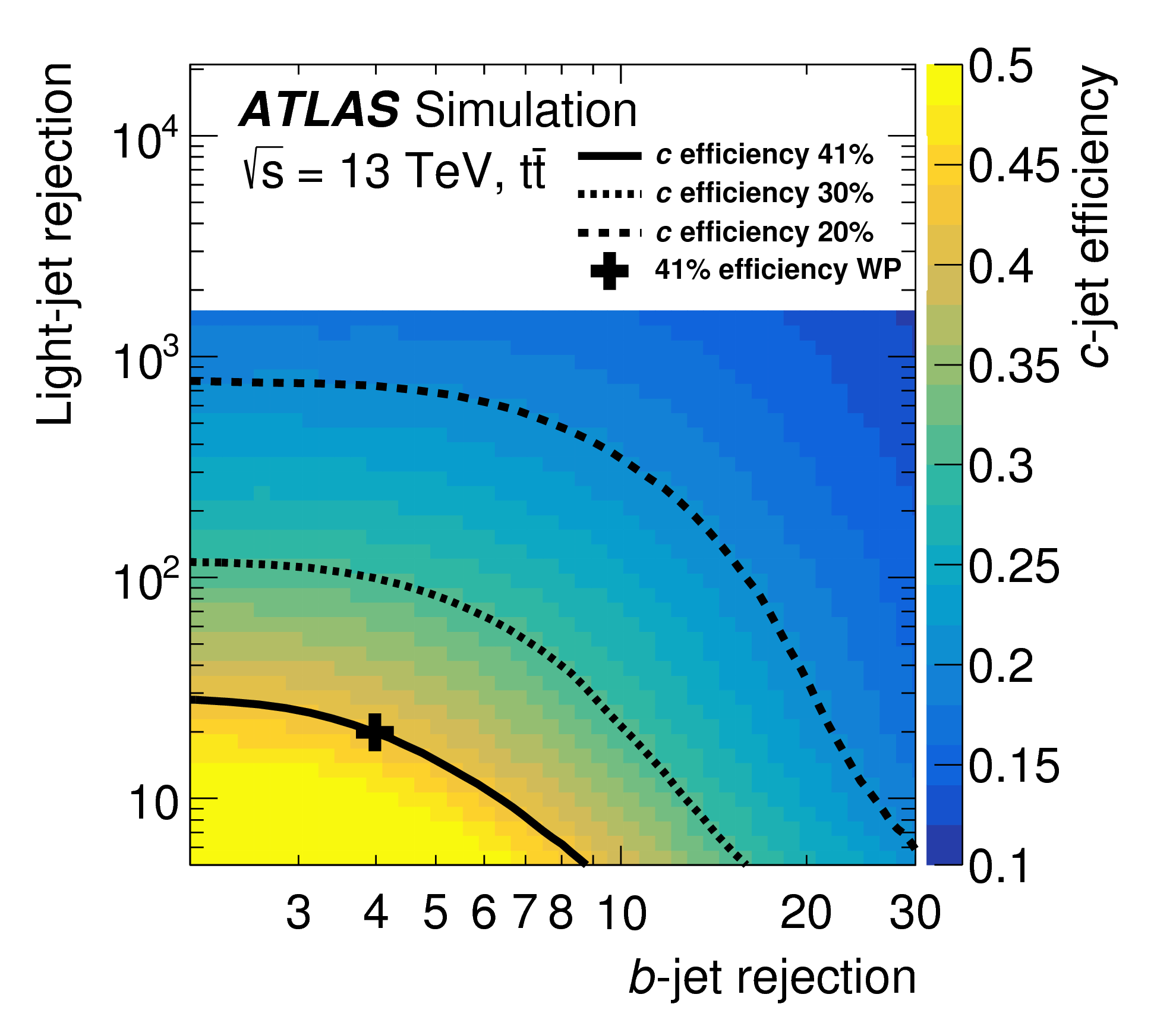}
\includegraphics[width=7.5cm]{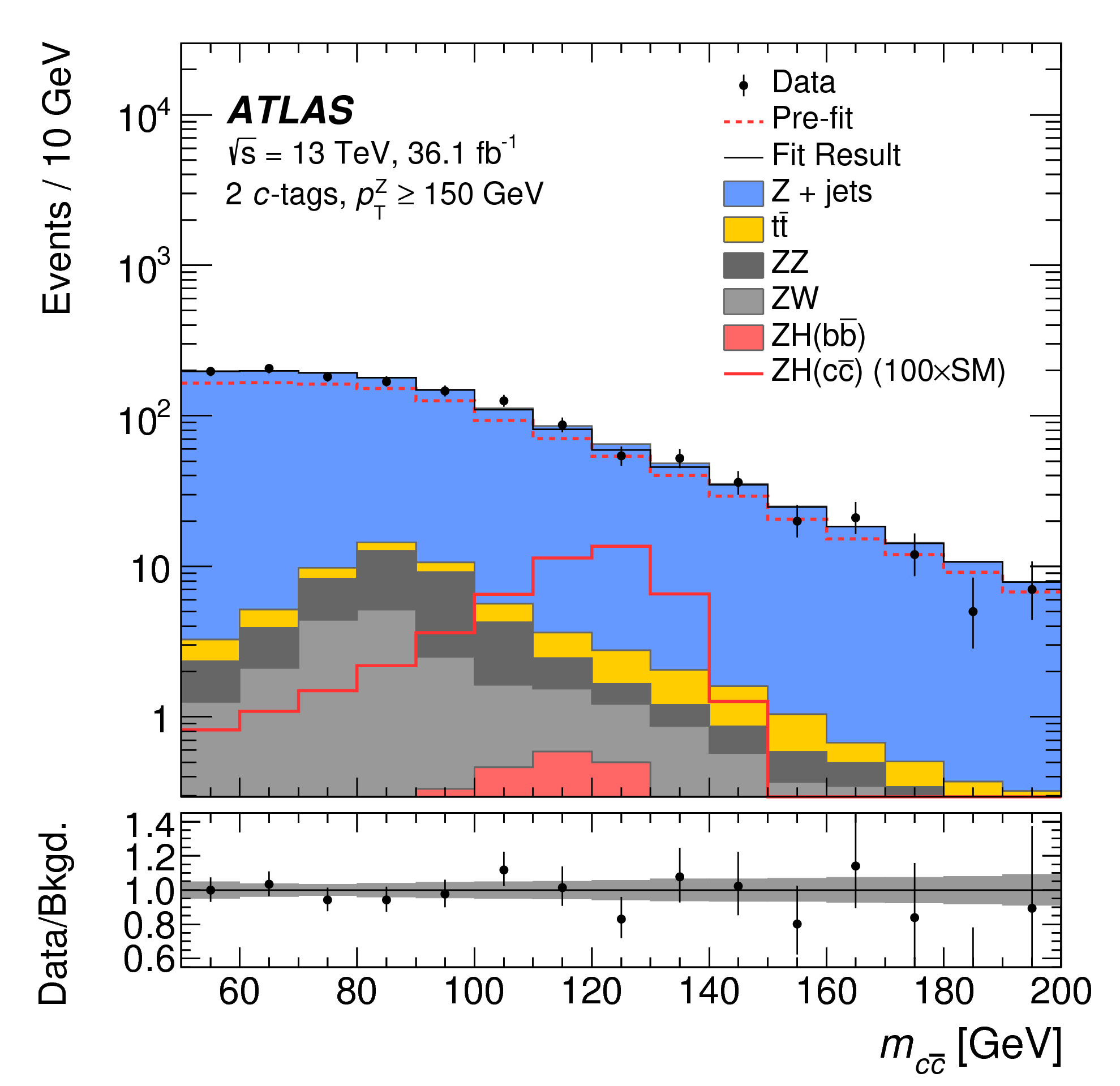}
\caption{The $c$-jet tagging efficiency as a function of the $b$-jet and light-jet (left). \mcc\ distribution in the events with \ptv $>$ 150 GeV (right)~\cite{VHcc}.}
\label{fig:VHccresult}
\end{figure}

\section{Search for the Higgs boson decay to dimuon}
A direct search for the SM decaying to dimuon using the data corresponding to an integrated luminosity of 36.1~\ifb is reported~\cite{Hmumu}. In the SM the branching ratio of dimuon is extremely small, 0.022\%. However, narrow dimuon mass peak in the final state can be observed around 125 GeV. The targeted production modes are gluon-fusion (ggF) and VBF which have first and second highest production cross section. Due to large Drell-Yan irreducible background, it is crucial to enhance signal-to-background ratio by the event categorization. After opposite-sign dimuon and low \MET and $b$-jet veto selections, a BDT based on the VBF topology is constructed to discriminate the VBF signal from the ggF and background. Figure~\ref{fig:Hmumuresult} shows a BDT distribution. The BDT distribution is split into three categories. Two VBF categories, VBF tight and VBF loose are defined from BDT score. The remaining low BDT events and events with less than two jets dominated by the ggF signal are further subdivided according to $p_{\mathrm{T}}^{\mu\mu}$ and muon $|\eta|$. To extract signal the $m_{\mu\mu}$ distributions of all signal regions are fit simultaneously. Figure~\ref{fig:Hmumuresult} shows $m_{\mu\mu}$ distribution in the tight VBF category. No significant excess is observed in this search. An observed (expected) upper limit on the signal strength at the 95\% confidence level is set to 3.0 (3.1). 

\begin{figure}[htb]
\centering
\includegraphics[width=7.5cm]{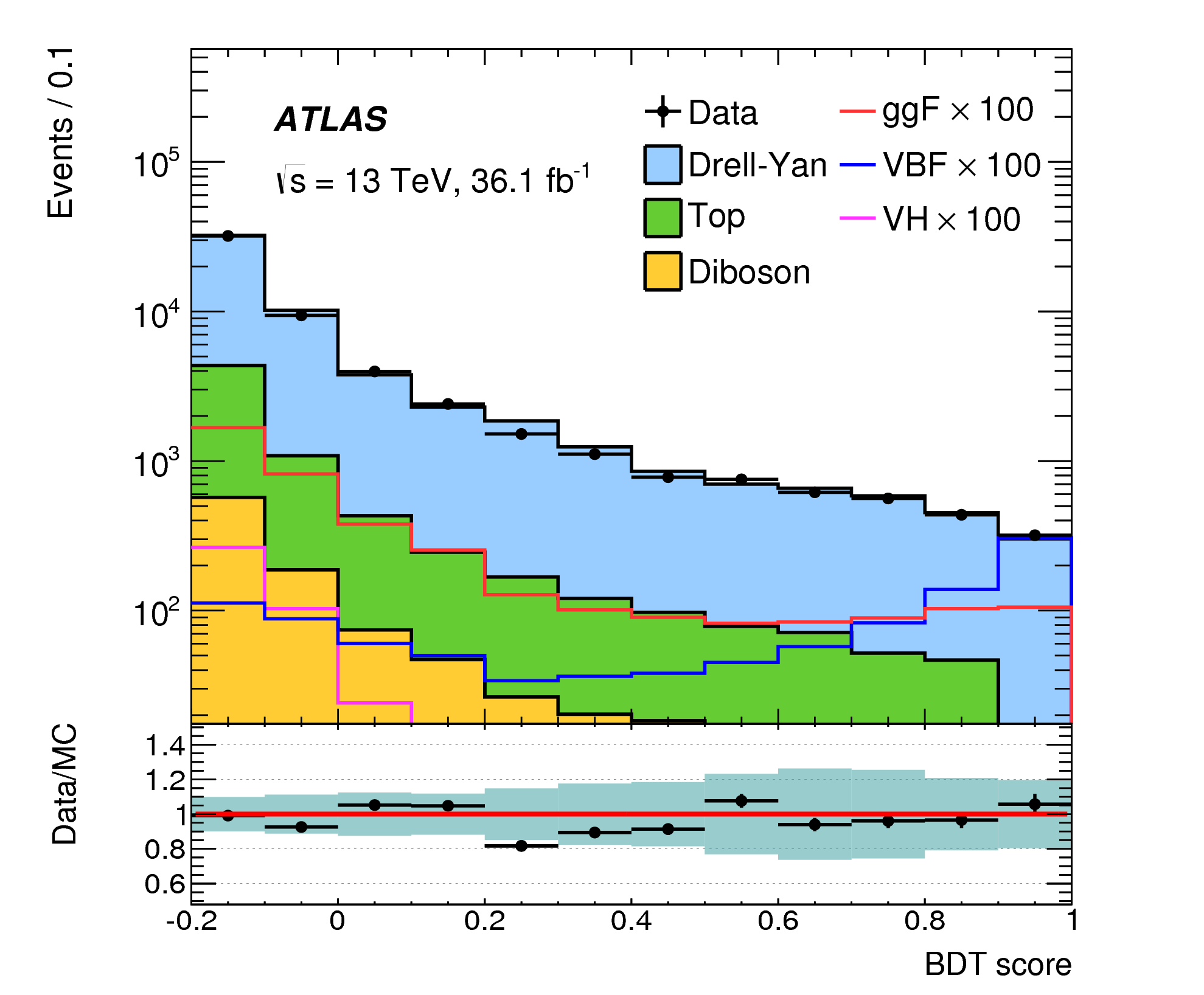}
\includegraphics[width=7.5cm]{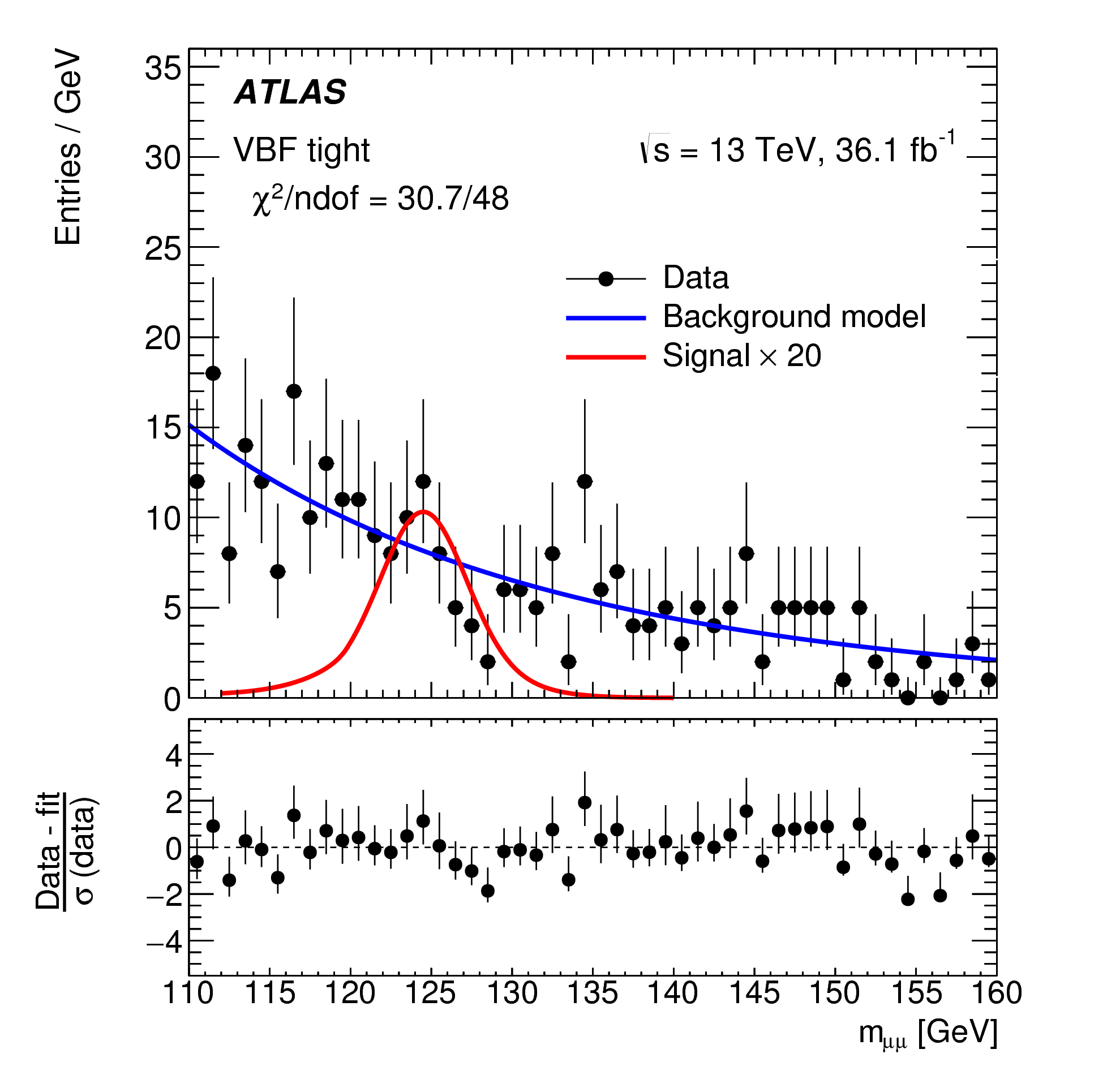}
\caption{The BDT output distribution (left). Dimuon mass distribution in the tight VBF category (right)~\cite{Hmumu}.}
\label{fig:Hmumuresult}
\end{figure}

\section{Conclusion}
The several decay modes of Higgs to two fermions are searched using ATLAS Run 2 dataset at a center-of mass energy of $\sqrt{s} = 13$ TeV in the proton-proton collisions produced by the LHC. Two searches for $H\rightarrow b\bar{b}$ are performed using the production in associated with vector boson and VBF. An excess over the expected background is observed with a significance of 3.6 standard deviations compared to an expectation of 4.0 in the combination with Run 1 and Run 2 dataset. The measured signal strength with respect to the SM expectation is found to be 0.90$\pm0.18(\textrm{stat.})^{+0.21}_{-0.19}(\textrm{syst.})$. In the search using VBF production, no significant excess is observed. 95\% confidence level upper limit on signal strength is set to 4.0, to be compared with expectation of $6.0^{+2.3}_{-1.7}$. 

Higgs decay to dimuon and $c\bar{c}$ (a pair of 2nd generation fermions) are searched. No significant excess is observed in both searches. The observed (expected) upper limit on signal strength at the 95\% confidence level is set to 3.0 (3.1) in the search for dimuon decay mode. An observed upper limit on $\sigma(pp\rightarrow ZH)\times B(H\rightarrow c\bar{c}$) is set to 2.7~pb, to be compared with an expectation of $3.9^{+2.1}_{-1.1}$~pb at the 95\% confidence level in the search for the $c\bar{c}$ decay mode.


\begin{thebibliography}{99}


\bibitem{ATLASdiscovery} 
  ATLAS Collaboration,
  Phys.\ Lett.\ B {\bf 716}, 1 (2012)

\bibitem{CMSdiscovery} 
  CMS Collaboration,
  Phys.\ Lett.\ B {\bf 716}, 30 (2012)

\bibitem{ATLASComb} 
  ATLAS Collaboration,
  Eur.\ Phys.\ J.\ C {\bf 76}, no. 1, 6 (2016)

\bibitem{ATLASSpinComb} 
  ATLAS Collaboration,
  Eur.\ Phys.\ J.\ C {\bf 75}, no. 10, 476 (2015)
  Erratum: [Eur.\ Phys.\ J.\ C {\bf 76}, no. 3, 152 (2016)]

\bibitem{ATLASMass} 
  ATLAS Collaboration,
  Phys.\ Lett.\ B {\bf 784}, 345 (2018)

\bibitem{ATLASCMSComb} 
  ATLAS and CMS Collaborations,
  JHEP {\bf 1608}, 045 (2016)

\bibitem{CMSttH} 
  CMS Collaboration,
  Phys.\ Rev.\ Lett.\  {\bf 120}, no. 23, 231801 (2018)

\bibitem{YR4} 
  LHC Higgs Cross Section Working Group,
  doi:10.23731/CYRM-2017-002
  arXiv:1610.07922 [hep-ph].

\bibitem{ATLAS} 
  ATLAS Collaboration,
  JINST {\bf 3}, S08003 (2008).


\bibitem{VHbb} 
  ATLAS Collaboration,
  JHEP {\bf 1712}, 024 (2017)

\bibitem{VHbbRun1} 
  ATLAS Collaboration,
  JHEP {\bf 1501}, 069 (2015)

\bibitem{VBFbb}
  ATLAS Collaboration,
  ATLAS-CONF-2016-063, \\
(https://atlas.web.cern.ch/Atlas/GROUPS/PHYSICS/CONFNOTES/ATLAS-CONF-2016-063/)

\bibitem{VHcc} 
  ATLAS Collaboration,
  Phys.\ Rev.\ Lett.\  {\bf 120}, no. 21, 211802 (2018)

\bibitem{Hmumu} 
  ATLAS Collaboration,
  Phys.\ Rev.\ Lett.\  {\bf 119}, no. 5, 051802 (2017)


\end{thebibliography}
\end{document}